\documentclass[11pt, oneside]{article}

\newcommand{\blind}{0}

\makeatletter
\renewcommand\section{\@startsection {section}{1}{\z@}%
	{-3.5ex \@plus -1ex \@minus -.2ex}%
	{2.3ex \@plus.2ex}%
	{\normalfont\fontfamily{phv}\fontsize{16}{19}\bfseries}}
\renewcommand\subsection{\@startsection{subsection}{2}{\z@}%
	{-3.25ex\@plus -1ex \@minus -.2ex}%
	{1.5ex \@plus .2ex}%
	{\normalfont\fontfamily{phv}\fontsize{14}{17}\bfseries}}
\renewcommand\subsubsection{\@startsection{subsubsection}{3}{\z@}%
	{-3.25ex\@plus -1ex \@minus -.2ex}%
	{1.5ex \@plus .2ex}%
	{\normalfont\normalsize\fontfamily{phv}\fontsize{14}{17}\selectfont}}
\makeatother

\usepackage{amsmath}
\usepackage{graphicx}
\usepackage{enumerate}
\usepackage[dvipsnames]{xcolor}
\usepackage{natbib} 
\usepackage{url} 
\usepackage{authblk}
\usepackage{fullpage}
\usepackage{amssymb, amsfonts}
\usepackage[utf8x]{inputenc}
\usepackage[T1]{fontenc}
\usepackage{siunitx}
\usepackage[version=3]{mhchem}
\usepackage{epsfig}
\usepackage{latexsym}
\usepackage{bm}
\usepackage{tabularx}
\usepackage{booktabs} 
\usepackage{enumerate}
\usepackage[titletoc]{appendix}
\usepackage{mathtools}
\usepackage{bbm}
\usepackage{multirow}
\usepackage{float} 
\usepackage{subcaption}
\usepackage[pdftitle={Title},
pdfauthor={Baik, Ko},
pdffitwindow=true,
breaklinks=true,
colorlinks=true,
urlcolor=blue,
linkcolor=Maroon,
citecolor=Maroon,
anchorcolor=red]{hyperref}

\graphicspath{{Figures/}}
\numberwithin{equation}{section}

\makeatletter
\def\ScaleWidthIfNeeded{%
	\ifdim\Gin@nat@width>\linewidth
	\linewidth
	\else
	\Gin@nat@width
	\fi
}
\def\ScaleHeightIfNeeded{%
	\ifdim\Gin@nat@height>0.9\textheight
	0.9\textheight
	\else
	\Gin@nat@width
	\fi
}
\makeatother
\setkeys{Gin}{width=\ScaleWidthIfNeeded,height=\ScaleHeightIfNeeded,keepaspectratio}%

\begin{document}
	
	\def\spacingset#1{\renewcommand{\baselinestretch}%
		{#1}\small\normalsize} \spacingset{1}
	
	\if0\blind
	{
		\title{\bf QoS-aware energy-efficient workload routing and server speed control policy in data centers: a robust queueing theoretic approach}
		\author{Seung Min Baik $^a$ and Young Myoung Ko $^a$ \\
			$^a$ Department of Industrial and Management Engineering, Pohang University of Science\\ and Technology, Pohang, Gyeongbuk, 37673, South Korea }
		\date{}
		\maketitle
	} \fi
	
	\if1\blind
	{
		
		\title{\bf \emph{IISE Transactions} \LaTeX \ Template}
		\author{Author information is purposely removed for double-blind review}
		
		\bigskip
		\bigskip
		\bigskip
		\begin{center}
			{\LARGE\bf \emph{IISE Transactions} \LaTeX \ Template}
		\end{center}
		\medskip
	} \fi
	\bigskip
	
	
	\begin{abstract}
		Operating cloud service infrastructures requires high energy efficiency while ensuring a satisfactory service level. Motivated by data centers, we consider a workload routing and server speed control policy applicable to the system operating under fluctuating demands. Dynamic control algorithms are generally more energy-efficient than static ones. However, they often require frequent information exchanges between routers and servers, making the data centers' management hesitate to deploy these algorithms. This study presents a static routing and server speed control policy that could achieve energy efficiency similar to a dynamic algorithm and eliminate the necessity of frequent communication among resources. We take a robust queueing theoretic approach to response time constraints for the quality of service (QoS) conditions. Each server is modeled as a $G/G/1$ processor sharing queue, and the concept of uncertainty sets defines the domain of stochastic primitives. We derive an approximative upper bound of sojourn times from uncertainty sets and develop an approximative sojourn time quantile estimation method for QoS. Numerical experiments confirm the proposed static policy offers competitive solutions compared with the dynamic algorithm.
	\end{abstract}
		
	\noindent%
	
	\spacingset{1.5}
	\renewcommand{\arraystretch}{0.7}
	
	\section{Introduction}
	\label{sec:introduction}
	
	The demand for computing power has been skyrocketing to process exploding data from mobile devices and associated cloud services. Global ICT companies such as Amazon, Google, and Facebook have been aggressively building large-scale data centers to fulfill the demand for computing power while pursuing energy efficiency for offering services with low carbon emissions. Currently, international ICT firms believe that data centers should contribute to improving environmental sustainability \citep{Holzle2020}, which is one of the most critical necessities for the survival of humanity. A recent report by Google \citep{Google2019} states that designing efficient data centers is one of its top priorities. Data centers consumed approximately 200 TWh of electricity in 2018, that is, approximately 1\% of the total electricity consumption worldwide in 2018 \citep{Jones2018}. Cisco, a global ICT company, expected that network traffic and workload in data centers will increase by 80\% and 50\%, respectively, by 2021; this implies a rapid increase in energy consumption \citep{Cisco2018}. However, current data center operations are inefficient, and average CPU utilization remains low at around 20\% \citep{Jiang2019, Cortez2017}. Energy efficiency can be significantly improved by reducing energy-inefficient operations. As servers and cooling systems account for more than 80\% of energy consumption in a data center \citep{Pelley2009, Rong2016}, we focus on server energy efficiency.
	
	The studies on energy efficiency in data centers approach the problem at two levels: planning and operation. Studies at the planning level focused on designing data centers through static resource allocation, such as server virtualization and server clustering. Once decided, modifying the setting in real time is not easy. \citet{Xiong2019} formulated an optimization model for server clustering and proposed a metaheuristics algorithm to solve the mixed-integer non-linear programming problem. \citet{Urgaonkar2007}, \citet{Gallego2013}, and \citet{Cho2018} also addressed the planning-level problem in different ways. Under the context established by the planning level, research at the operation level examined the optimal load balancing and server speed scaling to process real-time requests. Because of the real-time nature of operations, \textit{dynamic} load balancing and control algorithms have been considered in most of the energy efficiency studies at this level. \citet{cho2020stabilizing} studied performance stabilization in a processor sharing queue with a time-varying traffic pattern considering the mean virtual response time. \citet{George2001} and \citet{Ata2006} proposed control methods for single station systems based on quantitative mathematics such as Markov decision processes and queueing theory. \citet{Ko2014}, \citet{cho2022power}, and \citet{Andrew2009} proposed heuristic control methods. \citet{Bilal2015} developed a power-aware scheduling algorithm to simultaneously reduce makespan and energy consumption. Leading ICT companies, including Google and Amazon, are also utilizing artificial intelligence and machine learning techniques \citep{Google2019, Amazon2019}.
	
	Practically, however, implementing dynamic algorithms for a load balancer and servers is challenging. Dynamic algorithms, including distributed ones, often require frequent communications between network resources (e.g., routers and servers manufactured by different vendors and running heterogeneous operating systems to share their current status). Even if the network resources are equipped or retrofitted with a \textit{universal} software enabling all those communications, the management in charge of operation may be more concerned with the reliability of their service and reluctant to take a risk by adopting dynamic control algorithms. When we contacted a company operating several large data centers in Korea, the manager initially expressed interest but finally declined the use of dynamic algorithms for the aforementioned reason.
	
	In this regard, we investigate a static control policy that does not require frequent updates of the real-time status of resources; we determine the routing probabilities and server speeds once in a while, and resources autonomously run according to the policy. We use the terms \textit{control} and \textit{scaling} interchangeably throughout the study. Static control methods are generally less efficient than dynamic ones, particularly for situations with high variance and load fluctuations \citep{Chandra2009, Wierman2009}. \citet{Chen2015}, who investigated the interaction between load balancing and speed scaling, also pointed out that heterogeneity within a system is the main cause of inefficiency regardless of the number of servers.
	
	To overcome this drawback, we attempt to develop a static control policy that can compete with the dynamic algorithm in terms of performance. We also explore situations under which the static policy works well. The term \textit{static} does not indicate that the problem is solved only once and then used indefinitely. Depending on the changes in traffic patterns, operators can solve the problem again and update the routing probabilities and server speeds. The difference in the dynamic control is based on how often the control decisions are made: \textit{real time or not}.
	
	Previous studies at the operation level considering the quality of service (QoS) focused only on task scheduling (routing) or slack reclamation (speed scaling) \citep{Rizvandi2012, Uddin2015}. Various QoS metrics were handled in studies in which task scheduling and slack reclamation were considered simultaneously; however, most of these studies involved non-probabilistic hard constraints in the model \citep{Scarpiniti2018, Gul2016}. Probabilistic QoS constraints were handled in only a few studies; however, the QoS related to the response time was rarely considered \citep{Karpowicz2016}. For example, \citet{Baccarelli2017} proposed a dynamic network flow control (routing) and virtual machine frequency scaling algorithm. Their QoS is related to the server’s utility or processing capability from the operator’s perspective.
	
	However, \citet{Ko2014} employed the solution of an optimization problem for workload routing and server speed scaling while considering sojourn time-related probabilistic QoS constraints. Their algorithm successfully reflected the real-time status of resources using an iterative method converging to the optimal solution of the problem. One of the most crucial advantages of this approach was that it only required communications between the load balancer and each server, not inter-server communications, which enabled distributed control. Their dynamic algorithm satisfied the response time constraints regarding the service level agreements; however, the solution appeared significantly conservative as it used a loose upper bound for the constraints. For instance, as we will see later in Section \ref{sec:exp}, the average delay probability with given threshold 5 from the simulation is 0.000034 (i.e., $\mathbb{P}(S \geq 5) \approx 0.000034$) when the response time constraint is set to be $\mathbb{P}(S \geq 5) \leq 0.05$---delay time threshold 5 with violation probability 0.05.
	
	Therefore, we aim to derive a static policy with a tighter bound using robust queueing theory rather than the heavy-traffic approximation by \citet{Ko2014}. We describe the difference between the two approaches in detail and justify the reason for adopting a robust queueing theoretic approach in Section \ref{subsec:rqt}. Our static policy routes an incoming request to a server through a fixed probabilistic rule without real-time communications and operates the server at a constant speed when serving workload; a server runs at the minimum speed when idle.
	
	Based on this information, the contributions of this study can be summarized as follows:
	\begin{itemize}
		\item The proposed static control policy satisfies the response time constraints less conservatively with more competitive energy efficiency than the existing dynamic control algorithm. \nonumber
		\item This is the first study in which robust queueing theory is applied to processor sharing queues ($G/G/1/PS$ queues) suitable for modeling the CPU of a server using a time-slicing policy. 
		\item We introduce new uncertainty sets for the processor sharing (PS) discipline by extending previous results for the first-come-first-served (FCFS) queue \citep{Bandi2015, Whitt2018, Whitt2019}. Additionally, we propose a methodology for determining variability parameters in PS queues to match the size of uncertainty sets to the target probability. \nonumber
	\end{itemize}
	
	The remainder of this study is organized as follows. Section \ref{sec:probdesc} describes the system settings and the formulation of an optimization problem for obtaining the static control policy. Section \ref{sec:sla} briefly introduces the robust queueing theory and explains the steps for deriving a tractable inequality that approximates the response (sojourn) time constraint. Section \ref{sec:exp} presents numerical experiments conducted by comparing the performances of the proposed static policy and the dynamic control algorithm; it also presents a discussion on the results related to the static policy. Section \ref{sec:con} provides conclusions and suggestions for the future research.
	
	\section{Problem description}
	\label{sec:probdesc}
	
	Section \ref{subsec:systemdescription} describes the system with the probabilistic routing scheme. Section \ref{subsec:optmodel} constructs an optimization model that minimizes energy consumption under QoS conditions concerning response time.
	
	\subsection{\emph{System description}}
	\label{subsec:systemdescription}
	
\begin{figure}[t]
		\centering
		\includegraphics[width=0.6\linewidth]{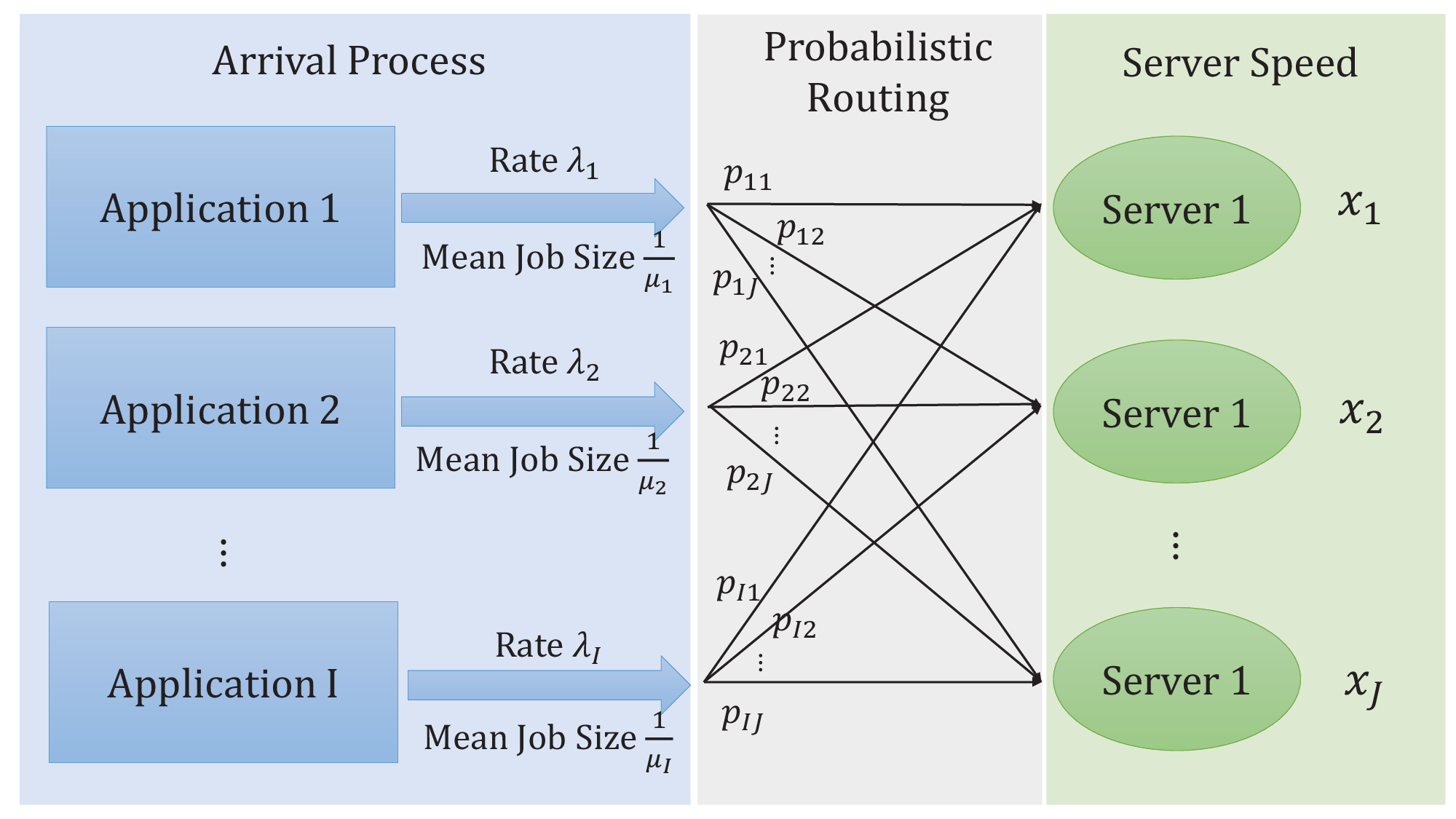}
		\caption{System model with probabilistic routing scheme}
		\label{figure:system}
	\end{figure}

	We consider a system consisting of $I$ applications and $J$ servers with a router, as shown in Figure~\ref{figure:system}. The individual servers can have different power functions and handle various applications. The server clusters and corresponding application assignments (virtualization) are given. Routing service requests follow a time-invariant probabilistic rule with the probability matrix $P=[p_{ij}]$. The arrival process of each application follows a renewal process, where inter-arrival time and workload random variables follow some general distributions having finite means and variances. When a service request of an application arrives, the router (or the load balancer) assigns the request to one of the $J$ servers according to the predefined probability. Each server processes all requests in a round-robin fashion at the same rate (i.e., we model each server as a PS queue). Thus, analyzing the thinning and superposition of independent renewal processes is necessary to deal with each server. Servers operate at different speeds when they are either busy or idle. When the server \(j\) is in a busy period, it runs at a predetermined speed $x_j$. Otherwise, it operates at minimum speed $x_j^{min}$. The QoS conditions are defined in terms of response time; the probability that the response time of a request--the time between arrival and service completion--exceeds a specific threshold must fall below a particular level.
	
	Our model is a simplified representation of a real system that comprises a single router and servers with a single power function. The model, however, can be extended to multi-tier architecture or multi-core processors by suitably combining multiple models and adjusting power functions; this can be explored in future work.
	
	\subsection{\emph{Optimization model}}
	\label{subsec:optmodel}
	
	The optimization model’s objective is to minimize servers’ energy consumption in a data center. The decision variables are routing probability matrix $P = [p_{ij}]$ and servers’ speed $x_j$. We write the optimization model denoted by (M1) as follows:
	\begin{align}
		\textrm{(M1)} \quad  \text{Min}\ & \sum_{j\in\mathcal{S}} \left(C_j(x_j)\rho_j  + C_j\left(x_j^{min}\right) (1-\rho_j)\right)  \nonumber \\
		\textrm{s.t.}\ & 0 \leq p_{ij} \leq 1, \quad \forall i \in \mathcal{A}, \forall j \in \mathcal{S}_i , \label{const:1} \\
		& \sum_{j \in \mathcal{S}_i} p_{ij} = 1, \quad \forall i \in \mathcal{A} , \label{const:2}\\
		&x^{min}_j\le x_j\le x^{max}_j, \quad \forall j\in\mathcal{S} , \label{const:3}\\
		&\mathbb{P} (S_j \geq \delta_j) \leq \varepsilon_j, \quad \forall j\in\mathcal{S}.\label{const:4}
	\end{align}
	
	In the objective function, $C_j(\cdot)$ denotes the power function of server \(j\). We consider long-run average power consumption under a static control policy. The fraction of the busy period for server $j$ equals traffic intensity {$\rho_j$} because the PS discipline satisfies work-conserving law. Thus, single server's average power consumption is a weighted sum of power function values with probabilities that the server is busy or idle. The objective value of the optimization model is the total power consumption over the entire server set denoted by $\mathcal{S}$. The power function is a convex function of server speed ($x_j$) (e.g., a cubic function in \citet{Wierman2012}). We use a more generic cubic representation, $C_j(x_j) = K_j + \alpha_j x_j^3$ as in \citet{Horvath2007}. The power functions do not have to be a polynomial function and can be any convex function. Changing the objective function, however, does not affect the crucial parts of our result---construction of uncertainty sets and approximation of QoS constraints.
	
	Constraint (\ref{const:1}) is a basic probability proposition. As shown in Figure~\ref{figure:system}, each arriving request of application $i$ is assigned to server $j$ with a probability of $p_{ij}$. $\mathcal{A}$ is the set of all applications, and $\mathcal{S}_i$ refers to the set of servers having application $i$. Constraint (\ref{const:2}) implies that the sum of the routing probabilities for each application should be one. Constraint (\ref{const:3}) describes the feasible speed range of the servers. Constraint (\ref{const:4}) is the direct representation of the QoS conditions, stating that the probability of the steady-state response time of server $j$, $S_j$, exceeding the threshold $\delta_j$ should be lower than the small probability of $\varepsilon_j$. Constraint~(\ref{const:4}) sets a lower bound of server speed (or an upper bound of the response time) and requires a queueing approximation for tractability. Making a bound as tight as possible is critical because a loose bound increases server speed and causes unnecessary energy waste. Therefore, the approximation of constraint~(\ref{const:4}) is an essential part of this study. 
	
	Several other significant sources consume electricity, such as storage, memory, and network, other than CPUs. However, combining all those sources into a single model highly complicates the problem. Therefore, we would restrict our focus on workload routing and server speed scaling for CPU-intensive jobs with the QoS conditions, which bound the tail probability of response time.
	
	\section{QoS constraints}
	\label{sec:sla}
	
	In this section, we reformulate the QoS conditions using robust queueing theory. Section~\ref{subsec:rqt} briefly explains the robust queueing theory. Section~\ref{subsec:fcfsps} describes how the robust queueing theoretic approach differs between FCFS and PS queues. Sections~\ref{subsec:us}--\ref{subsec:main_variability_parameters} describe the steps for deriving a tractable condition that approximates constraint (\ref{const:4}). Section~\ref{subsec:us} constructs uncertainty sets suitable for PS queues, Section~\ref{subsec:rtb} derives the response time bound regarding the uncertainty sets, and Section~\ref{subsec:main_variability_parameters} explains a procedure for choosing appropriate variability parameters. Sections~\ref{subsec:thinning} and~\ref{subsec:superposition} describe how the uncertainty sets vary for the thinning and superposition of arrival processes. Section \ref{subsec:serverwise} presents the analysis result for each server.
	
	\subsection{\emph{Robust queueing theory}}
	\label{subsec:rqt}
	
	\citet{Bandi2015} developed the robust queueing theory as an alternative approach to queueing analysis based on robust optimization. Traditional queueing theory has limitations such that even simple queueing systems are not tractable when the arrival processes or workloads are non-Markovian. To avoid this difficulty, the analysis of non-Markovian queues relies on simulation and approximation.
	
	A large number of simulation runs can generate accurate results. However, each run requires a long time to produce an outcome, particularly in heavy-tailed systems. Approximation approaches can provide explicit mathematical expressions though they work well in some limited situations. Our study takes the approximation approach because we require analytical expressions to be embedded into the optimization model. A popular method, known as the heavy-traffic approximation, tends to underestimate the response time in our prestudy experiments, which causes violations of the QoS conditions. Therefore, we take a new approach---robust queueing theory---to construct robust and tractable expressions for constraint (\ref{const:4}) while utilizing the findings of heavy-traffic approximation results; we take the functional form of the heavy-traffic approximation for some parameters, similar to \citet{Bandi2015} that uses the functional form of the Kingman's bound \citep{kingman1970inequalities}. 
	
	The fundamental difference between the heavy-traffic approximation and robust queueing theory lies in the treatment of random variables. The heavy-traffic approximation first derives the distribution of the target performance measure, such as response time. Then, it calculates the quantile of the random variable using a cumulative distribution function. The robust queueing theory reverses this order. We compare the robust queueing theoretic approach to the heavy traffic approximation used by \citet{Ko2014} that approximates the response time process as a reflected Brownian motion using the steady-state results for the $G/G/1/PS$ queue in \cite{Zhang2008}:
	\begin{equation}
		S \approx \tilde{V}(\infty) \approx \beta \tilde{Z}(\infty), \nonumber
	\end{equation}
	where $\tilde{V}$ is a virtual response time and $\tilde{Z}$ is a reflected Brownian motion with drift $\theta/\beta_e$ and variance $\beta(C_a^2 + C_s^2)/\beta_e^2$. We refer the readers to Section 3 in \citet{Ko2014} for a detailed explanation of other parameters. The approximate probability of exceeding a threshold, $\delta$, is given by:
	\begin{equation}
		\mathbb{P}(S \geq \delta) \approx \exp\left(-\frac{\theta(1+C_s^2)\delta}{\beta\left(C_a^2 + C_s^2\right)}\right). \nonumber
	\end{equation}
	The QoS constant $\varepsilon$ is treated at the last moment.
	
	In the robust queueing theoretic approach, however, $\varepsilon$ appears at the beginning of the analysis. We first restrict the probabilistic region for stochastic primitives and then proceed with the analysis. For better understanding, we defer the precise mathematical details about the definition of uncertainty sets' probability and arrival index on purpose to Sections~\ref{subsec:fcfsps}--\ref{subsec:rtb}. We only depict an abstract concept here. Suppose we define the uncertainty sets of stochastic primitives, 
	inter-arrival times $\mathbf{T}{(n) = (T_1, T_2, \dots, T_n)}$, and workloads $\mathbf{X}{(n) = (X_1, X_2, \dots, X_n)}$, with probability $1-\varepsilon$ as follows:
	\begin{equation}
		\mathbb{P}\left(\mathbf{T}{(n)} \in \mathcal{U}^a{(n)}, \mathbf{X}{(n)} \in \mathcal{U}^s{(n)}\right) = 1-\varepsilon. \nonumber
	\end{equation}
	We derive an approximative upper bound of the worst-case sojourn time among the uncertainty sets of {$\mathbf{T}(n)$} and {$\mathbf{X}(n)$} as follows: 
	\begin{equation*}
		\underset{\mathbf{T}{(n)} \in \mathcal{U}^a{(n)}, \mathbf{X}{(n)} \in \mathcal{U}^s{(n)}}{\max}S_n (\mathbf{T}{(n)}, \mathbf{X}{(n)}) \leq {S_{\text{UB}}}, \ \forall n,
		\nonumber
	\end{equation*}
	and use it to construct a tractable condition that can replace the QoS constraint, that is, we substitute $\mathbb{P}(S \geq \delta)$ with ${S_{\text{UB}}} \leq \delta$.	Considering \(\varepsilon\) in advance is essential in constructing uncertainty sets. Adopting a robust queueing theory can provide analytical representations with reasonable accuracy and ensure that the QoS conditions are satisfied not too conservatively.
	
	\subsection{\emph{Comparison of FCFS and PS queues}}
	\label{subsec:fcfsps}
	
	Before explaining the derivation of the worst-case sojourn time and its bound, we review the robust queueing analysis for FCFS queues and determine what we should modify under PS discipline. Previous studies on robust queueing theory \citep{Bandi2015, Whitt2018, Whitt2019} follow a similar procedure for response time analysis. We can break it down into the following four steps:
	\begin{enumerate}
		\item Construct uncertainty sets $\mathcal{U}^a{(n)}$ and $\mathcal{U}^s{(n)}$ for $n$ inter-arrival time and workload random variables $\mathbf{T}{(n) = (T_1, T_2, \dots, T_n)}$ and $\mathbf{X}{(n) = (X_1, X_2, \dots, X_n)}$.\label{step:1}
		\item Express the exact response time of the $n^{\text{th}}$ arriving job, $S_n(\mathbf{T}{(n)}, \mathbf{X}{(n)})$, via stochastic primitives $\mathbf{T}{(n)}$ and $\mathbf{X}{(n)}$.\label{step:2}
		\item Calculate the worst-case response time of the $n^{\text{th}}$ arriving job with respect to the stochastic primitives within the uncertainty sets, $\hat{S}_n(\mathcal{U}^a{(n)}, \mathcal{U}^s{(n)}) = \underset{\mathbf{T}{(n)} \in \mathcal{U}^a{(n)}, \mathbf{X}{(n)} \in \mathcal{U}^s{(n)}}{\max}S_n (\mathbf{T}{(n)}, \mathbf{X}{(n)})$. \label{step:3}
		\item Derive a bound 
		$S_{\text{UB}}$ that is independent of \(n\) and is an upper bound of $\hat{S}_n(\mathcal{U}^a{(n)}, \mathcal{U}^s{(n)})$ for all $n$.\label{step:4}
	\end{enumerate}
	In steps \ref{step:1} to \ref{step:3}, the target performance measure is bound by the desired probability; however, the bound is valid only for the request with a specific (finite) arrival index. Thus, step \ref{step:4} generalizes the result for the arbitrary arrival index and, consequently, culminates in the steady-state response time. 
	
	However, we cannot directly apply the above approach to a PS queue. For a $G/G/1/FCFS$ queue, we can represent the exact expression of the response time in step \ref{step:2} using a linear form of stochastic primitives. However, for a $G/G/1/PS$ queue, the expression of the PDE-form is only available. Equations (\ref{eqn:fcfs}) and (\ref{eqn:ps}) represent the response time expressions in the $G/G/1/FCFS$ and $G/G/1/PS$ queues, respectively. For a FCFS queue, by the Lindley recursion \citep{Lindley1952},
	\begin{equation}
		S_n^{FCFS} = \underset{1 \leq k \leq n}{\max} \left(\sum_{i=k}^n X_i - \sum_{i=k}^{n-1} T_i\right), \label{eqn:fcfs}
	\end{equation}
	where $S_n^{FCFS}$, $X_n$, and $T_n$ denote the response time of the $n^{\text{th}}$ arriving job under the FCFS discipline, the workload of the $n^{\text{th}}$ arriving job, and the inter-arrival time between the $n^{\text{th}}$ and the $(n+1)^{\text{th}}$ arrivals, respectively. For a PS queue, we have
	\begin{equation}
		\begin{gathered}
			A_n = \sum_{i=1}^{n} T_i, \quad D_n = A_n + S_n^{PS}, \quad X_n = \int_{A_n}^{D_n} \frac{1}{N(t)}dt ,\\ N(t) = \sum_{n=1}^{\infty} \mathbbm{1}_{ \{A_n \leq t < D_n\} } =  \sum_{n=1}^{\infty} \mathbbm{1}_{ \{A_n \leq t\} } -  \sum_{n=1}^{\infty} \mathbbm{1}_{\{D_n \leq t \}}
		\end{gathered}\label{eqn:ps}
	\end{equation}
	where $S_n^{PS}$, $N(t)$, $A_n$, and $D_n$ denote the response time of the $n^{\text{th}}$ arriving job under the PS discipline, the number of unfinished jobs in the queue at time $t$, and the arrival and departure time of the $n^{\text{th}}$ arriving job, respectively. The rate at which the remaining workload of each job decreases is inversely proportional to $N(t)$ for a PS queue. We cannot express the response time in a linear form of $X_n$ and $T_n$.
	
	Therefore, we cannot directly follow the same steps \ref{step:1} to \ref{step:4} to analyze a PS queue. The optimization problem is almost intractable if we embed equation~(\ref{eqn:ps}) as a model constraint. To avoid the PDE-form, we integrate steps \ref{step:2} and \ref{step:3} in the robust queueing analysis procedure by creating a loose worst-case upper bound instead of seeking an alternative near-exact expression or a heuristic solution. Although our approach loosens the inequality and produces more conservative results, the experimental results in Section \ref{sec:exp} demonstrate that our method successfully guarantees constraint~(\ref{const:4}) without being overly conservative.
	
	\subsection{\emph{Uncertainty sets}}
	\label{subsec:us}
	
	We adopt a similar approach to \citet{Bandi2015} and \citet{Whitt2018} in constructing uncertainty sets. Here, we apply the central limit theorem (CLT), that is, for independent and identically distributed (i.i.d.) random variables $\xi_i$ with the same distribution as $\xi$, the quantity $(\sum_{i=1}^n \xi_i - n\mathbb{E}[\xi])/\sqrt{n\text{Var}[\xi]}$ becomes asymptotically standard normal as $n$ approaches infinity. To apply the CLT, this study assumes the random variables that characterize the arrival process (i.e., inter-arrival time and workload) to have a finite mean and variance, which is a reasonable assumption for practical purposes. The worst-case response time is achieved when the inter-arrival times are short and the workloads are large. Therefore, we construct the uncertainty sets with one-sided inequalities to consider situations with shorter inter-arrival times and larger workloads. 
	
	We construct the uncertainty set of inter-arrival times, $\mathcal{U}^a(n)$ as follows:
	\begin{equation}
		\mathcal{U}^a(n) = \left\{\mathbf{T}(n) = (T_1, T_2, \dots, T_n) \left| \frac{\sum_{i=k}^n T_i - \frac{n-k+1}{\lambda}}{\sqrt{n-k+1}} \geq -\Gamma_a, \ \forall k, 1 \leq k \leq n \right\}. \right.  \label{eqn:ua}
	\end{equation}
	Here, $1/\lambda$ and $\sigma_a$ are the mean $\mathbb{E}[T_i]$ and standard deviation $\sqrt{\text{Var}[T_i]}$ of the inter-arrival times, respectively, and $\Gamma_a$ is a constant called variability parameter related to the size of the uncertainty set or its corresponding probability. Similarly, we construct the uncertainty set of service workloads, $\mathcal{U}^s(n)$, as follows:
	\begin{equation}
		\begin{aligned}
			\mathcal{U}^s(n) = \left\{\mathbf{X}(n) = (X_1, X_2, \dots, X_n) \left| \frac{\sum_{i=k}^n X_i - \frac{n-k+1}{\mu}}{\sqrt{n-k+1}} \leq \Gamma_s, \ \forall k, 1 \leq k \leq n \right. \right. \\
			\textrm{ and}  \left.\ \frac{\sum_{i=k}^{n-1} X_i - \frac{n-k}{\mu}}{\sqrt{n-k}} \leq \Gamma_s, \ \forall k, 1 \leq k \leq n-1 \right\}. 
		\end{aligned}
		\label{eqn:us}
	\end{equation}
	Here, $1/\mu$ and $\sigma_s$ are the mean $\mathbb{E}[X_i]$ and standard deviation $\sqrt{\text{Var}[X_i]}$ of the workloads, respectively. $\Gamma_s$ is a variability parameter for the uncertainty set of service workloads.
	
	The uncertainty set for service workloads {$\mathcal{U}^s(n)$} differs from the one defined by \citet{Bandi2015}. The sojourn time can be expressed using the terms of finite service workload sums, $\sum_{i=k}^n X_i$, in an FCFS queue. The sojourn time in a PS queue, however, cannot be explicitly expressed with stochastic primitives; it can only be bounded via terms of $\sum_{i=k}^n X_i + \sum_{i=k}^{n-1} X_i$, and the derivation procedure is explained in detail in Section 1.1 in the online supplement. Thus, the modified uncertainty set structure bounds above both $\sum_{i=k}^n X_i$ and $\sum_{i=k}^{n-1} X_i$. 
	
	\subsection{\emph{Response time bound}}
	\label{subsec:rtb}
	
	This section derives the bound of the response time from the uncertainty sets defined in equations (\ref{eqn:ua}) and (\ref{eqn:us}). Here, we assume that variability parameters $\Gamma_a$ and $\Gamma_s$ are predetermined to appropriate values such that the size of uncertainty sets corresponds to the target probability level $1-\varepsilon$. Finding these \(\Gamma_a\) and \(\Gamma_s\) is explained in the following section. Unlike FCFS queues, we cannot express the response time and its worst-case value within the uncertainty sets in an exact form of stochastic primitives for PS queues. Therefore, our methodology constructs an inequality for response time and directly obtains its upper bound. This procedure {merges} steps~\ref{step:2} and~\ref{step:3} in Section \ref{subsec:fcfsps}.
	
	Based on Theorem $3.1$ of \citet{Brandt2006}, the following holds for the steady-state sojourn (response) time random variables, $S^{PS}$ and $S^{FCFS}$, of $G/G/1/PS$ and $G/G/1/FCFS$ systems, respectively, that share the same stochastic primitives:
	\begin{equation}
		\mathbb{E} [S^{PS}] + \mathbb{E} [X] \leq 2 \mathbb{E} [S^{FCFS}]. \nonumber 
	\end{equation}
	Because we are dealing with a probabilistic QoS constraint, $\mathbb{P}(S \geq \delta) \leq \varepsilon$, in \textit{steady state}, we use the following inequality as an approximation:
	\begin{equation}
		S_{{n}}^{PS} + X_{n} \leq 2 S_{n}^{FCFS}, \label{eqn:approxineq}
	\end{equation}
	where \(S_{{n}}^{FCFS}\) and \(S_{n}^{PS}\) are the sojourn times of the $n^{\text{th}}$ arriving job under FCFS and PS disciplines, respectively. For a large enough $n$, we can think the queue is in steady state. Assuming this inequality, the worst-case sojourn time with respect to the stochastic primitives within uncertainty sets is bounded as follows (See Section 1.1 in the online supplement for the proof):
	\begin{equation}
		{\hat{S}_n^{PS}(\mathcal{U}^a(n), \mathcal{U}^s(n)) = \underset{\mathbf{T}(n) \in \mathcal{U}^a(n), \mathbf{X}(n) \in \mathcal{U}^s(n)}{\max}S_n^{PS} (\mathbf{T}(n), \mathbf{X}(n)) \leq S_{\text{UB}} := \frac{(\Gamma_a + \Gamma_s)^2 \lambda}{2(1-\rho)} + \frac{2-\rho}{\lambda}.} \label{eqn:sub}
	\end{equation}
	$S_{\text{UB}}$ is a valid upper bound regardless of $n$, which only relies on the characteristics of the arrival process and the size of the uncertainty sets. We note that the derivation draws on approximation; the selection of variability parameters approximately regulates the tightness of the bound.
	
	Section~\ref{subsec:main_variability_parameters} describes a procedure to find appropriate parameter values so that the following inequality about steady-state sojourn time random variable $S^{PS}$ (we drop the superscript PS now for notational convenience) holds not too conservatively:
	\begin{equation*}
		\mathbb{P}\left(S^{PS} \leq \frac{(\Gamma_a + \Gamma_s)^2 \lambda}{2(1-\rho)} + \frac{2-\rho}{\lambda}\right) \geq 1- \varepsilon.
	\end{equation*}		
	Therefore, we replace the QoS constraint with the following condition to deduce a heuristic policy:
	\begin{equation}
		\begin{aligned} 
			&S_{\text{UB}} = \frac{(\Gamma_a + \Gamma_s)^2 \lambda}{2(1-\rho)} + \frac{2-\rho}{\lambda} \leq \delta.
		\end{aligned} \label{eqn:slatoineq}
	\end{equation}
	Given an arrival process, server speed scaling only affects workload-related parameters $\Gamma_s$, $\mu$, or indirectly, $\rho$, in the aforementioned inequality.
	
	The variability parameter values bridge the uncertainty sets and the associated probability. \citet{Bandi2015} presented the procedure for determining the values as \textit{``translating the stochastic primitive data into robust primitive data, namely uncertainty sets with appropriate variability parameters.''} The following section describes our approach (i.e., such a procedure for PS queues) in detail. We only state here that the variability parameters do not have to rely on $n$ because we consider a large $n$ to approximate steady state.
	
	\subsection{\emph{Variability parameters}}
	\label{subsec:main_variability_parameters}
	
	This section describes how we determine the variability parameters to match the size of uncertainty sets with the target probability level under the PS discipline. The approximative upper bound $S_{\text{UB}}$ derived in Section~\ref{subsec:rtb} is calculated from parameters of an arrival process ($\mu, \lambda, \sigma_a$, $\sigma_s$) and variability parameters ($\Gamma_a$, $\Gamma_s$). Thus, given the arrival process, selecting the appropriate variability parameters is crucial for meaningful robust queueing analysis. Previous works \citep{Bandi2015, Whitt2018, Whitt2019} also highlighted such importance and provided different methods for determining the variability parameter values. Among them, we adopt a similar approach to \citet{Bandi2015}. They designed a functional form of variability parameters by considering the qualitative insights given by Kingman's bound \citep{kingman1970inequalities} and fitted the function coefficients using the simulation. Likewise, we follow a similar procedure while utilizing the findings of heavy-traffic approximation results. For this, it is important to ensure that the result of the robust queueing theoretic approach shares the same qualitative insights as the result of the traditional probabilistic framework.
	
	In this work, we utilize the tail probability approximation of \citet{Zhang2008} for the steady-state sojourn time in a PS queue as follows:
	\begin{equation*}
		\mathbb{P}(S \geq \delta) \approx \exp \left(-\frac{1+C^2_s}{C^2_a + C^2_s} \frac{1-\rho}{\rho} \lambda \delta\right) \leq \exp \left(-\frac{1}{C^2_a + C^2_s} \frac{1-\rho}{\rho} \lambda \delta\right).
	\end{equation*}
	Motivated by this approximation, we derive an approximative bound on the sojourn time quantile concerning the target probability level $1-\varepsilon$ as follows:
	\begin{equation}
		\delta \leq - \left(C^2_a + C^2_s\right) \frac{\rho}{(1-\rho)\lambda} \log \mathbb{P}(S \geq \delta) = \frac{\left(\rho \sigma_a^2 + \frac{\sigma^2_s}{\rho}\right)\lambda}{1-\rho} \left( -\log \varepsilon \right).
		\label{eq:approximate_st_bound_ht}
	\end{equation}
	Correspondingly, we consider the functional form of the variability parameters \(\Gamma_a\) and \(\Gamma_s\) as follows:
	\begin{equation}
		\begin{aligned}
			\Gamma_a &= f_a(\mu, \lambda, \sigma_a, \sigma_s; 1-\varepsilon) := \sigma_a,\\
			\Gamma_s &= f_s(\mu, \lambda, \sigma_a, \sigma_s; 1-\varepsilon) := \left(\theta_0 + \theta_1 \rho \sigma_a^2 + \theta_2 \frac{\sigma_s^2}{\rho} + \theta_3 \frac{ (1-\rho)(2-\rho)}{\lambda^2}\right)^{1/2} - \sigma_a,
		\end{aligned}
		\label{eqn:functional_form_vp_PS}
	\end{equation}
	where the traffic intensity is $\rho = \lambda/\mu$.
	This enables the approximative sojourn time upper bound, $S_{\text{UB}}$, to have functional dependence on $\lambda/(1-\rho)$ and $\rho\sigma_a^2, \sigma_s^2/\rho$ as the bound in equation~\eqref{eq:approximate_st_bound_ht} as follows:
	\begin{equation}
		S_{\text{UB}} =  \frac{(\Gamma_a + \Gamma_s)^2 \lambda}{2(1-\rho)} + \frac{2-\rho}{\lambda} 	= \frac{\left(\theta_1 \rho \sigma_a^2 + \theta_2 \frac{\sigma_s^2}{\rho}\right)\lambda}{2(1-\rho)} + \left( \frac{\theta_0 \lambda}{2(1-\rho)} + \left(\frac{\theta_3}{2} + 1\right)\frac{2-\rho}{\lambda} \right).
		\nonumber
	\end{equation}
	
	This is an analogue to the method for FCFS queues in \citet{Bandi2015}, and we explain it briefly here for comparison. They utilize the Kingman's bound \citep{kingman1970inequalities} on the expected steady-state waiting time for $G/G/1/FCFS$ queues, $\mathbb{E}\left[W^{FCFS}\right]$, to bound the expected steady-state sojourn time as follows:
	\begin{equation*}
		\mathbb{E}\left[S^{FCFS}\right] = \mathbb{E}\left[W^{FCFS} + X\right] = \mathbb{E}\left[W^{FCFS}\right] + \mathbb{E}\left[X\right] \leq \frac{(\sigma_a^2 + \sigma_s^2)\lambda}{2(1-\rho)} + \frac{1}{\mu}.
	\end{equation*}
	They consider the functional form of the variability parameters as follows:
	\begin{equation}
		\begin{aligned}
			\Gamma^{FCFS}_a &= f_a^{FCFS}(\rho, \sigma_a, \sigma_s) := \sigma_a, \\
			\Gamma^{FCFS}_s &= f_s^{FCFS}(\rho, \sigma_a, \sigma_s) := (\theta_0 + \theta_1 \sigma_s^2 + \theta_2 \sigma_a^2 \rho^2)^{1/2} - \sigma_a,
		\end{aligned}
		\label{eqn:bandi_variability_parameter_functions}
	\end{equation}
	and state that the upper bound of the worst-case sojourn time by robust queueing theory (From Theorem $2$) given by:
	\begin{equation*}
		\frac{(\Gamma_a^{FCFS} + \Gamma_s^{FCFS})^2 \lambda}{4(1-\rho)} + \frac{1}{\lambda} = \frac{(\theta_1 \sigma_s^2 + \theta_2 \sigma_a^2 \rho^2)\lambda}{4(1-\rho)} + \left(\frac{\theta_0 \lambda}{4(1-\rho)} +  \frac{1}{\lambda} \right)
	\end{equation*}
	has same functional dependence on $\lambda/(1-\rho)$ and $\sigma_a^2, \sigma_s^2$ as Kingman's bound. They mention that their \textit{``robust queueing theoretic approach leads to the same qualitative conclusions as stochastic queueing theory''}.
	
	As done in \citet{Bandi2015}, we first run simulations for a single server queue with different parameters of arrival process ($\lambda, \mu, \sigma_a, \sigma_s$) and calculate the sojourn time quantile with predefined probability level $1-\varepsilon$. Because we generate instances with a large number of arrivals and services to be close to the steady state (we track $S_n$ with sufficiently large $n$), the variability parameters do not have to consider index $n$. Then, we fit a function for the variability parameters in equation~\eqref{eqn:functional_form_vp_PS} such that the $S_{\text{UB}}$ value is adapted to a simulation-based quantile value. We present the detailed procedure of the numerical experiments and results in Section 2 in the online supplement.
	
	Once obtaining variability parameter values, we can substitute the QoS constraint with its approximate but tractable condition as in equation~\eqref{eqn:slatoineq}. However, the actual arrival process in our model is constructed through thinning and superposition of the external (original) arrival processes. The external arrivals of each application are probabilistically split or thinned by routing rules (thinning). This thinned process generates a flow of application $i$ to server $j$, and a particular server receives an aggregate incoming request from the various applications (superposition). Therefore, to properly handle server-wise QoS constraints, we must examine the relationship between uncertainty sets of external, thinned, and aggregated processes. Section~\ref{subsec:thinning} presents a discussion on the thinning of arrival process, and Section~\ref{subsec:superposition} explains the superposition of arrival processes.
	
	\subsection{\emph{Thinning of arrival process}}
	\label{subsec:thinning}
	
	\begin{table}[t]
		\caption{Notations for thinning of arrival process}
		\label{tab:not_thinning}
		\centering
		\scriptsize
		\begin{tabular}{cc}
			\toprule
			Symbol & Description\\
			\midrule
			$\lambda_i$ & Arrival rate of application $i$\\
			$\sigma_{a,i}$ & Standard deviation of the inter-arrival time of application $i$\\
			$\lambda_{ij}$ & Arrival rate of application $i$ into server $j$\\
			$\sigma_{a,ij}$ & Standard deviation of the inter-arrival time of application $i$ into server $j$\\
			$1/\mu_i$ & Average workload of application $i$\\
			$\sigma_{s,i}$ & Standard deviation of the workload of application $i$\\
			$P = [p_{ij}]$ & Routing probability matrix\\
			\bottomrule
		\end{tabular}
	\end{table}

	First, we investigate a thinned arrival process of application $i$ entering server $j$. We could use the thinning operator in Theorem 6 in \citet{Bandi2015} as it provides a good approximation result of the probabilistic routing policy. However, if the routing probability $p_{ij}$ is small, it might lead to an overly optimistic result. Thus, we choose to construct the uncertainty sets of the thinned arrival process by directly deriving the mean and variance of its inter-arrival time and service workload random variables.
	
	The server $j$ can take the request of application $i$ with probability $p_{ij}$. Let $T_{ij}$ be a random variable denoting the inter-arrival time of application $i$ to server $j$. We express $T_{ij}$ using the random variable denoting the inter-arrival time of application $i$ to the system, $T_i$, as follows:
	\begin{equation}
		\begin{aligned}
			\nonumber
			T_{ij} = \sum_{n=1}^{N_{ij}} T_i^n = 
			\begin{cases}
				T_i^1 & \text{if } N_{ij} = 1 \text{ with } \mathbb{P}(N_{ij}=1) = p_{ij} \ \\
				T_i^1 + T_i^2 & \text{if } N_{ij} = 2 \text{ with } \mathbb{P}(N_{ij}=2) = (1-p_{ij})p_{ij} \ \\
				\vdots
			\end{cases}
			,
		\end{aligned}
	\end{equation}
	where $T_i^n$'s are i.i.d. random variables with the same distribution as $T_i$ and $N_{ij}$ is a random variable denoting the number of external arrivals of application $i$ until one application $i$ arrival is routed to server $j$ for the first time, i.e., $N_{ij}$ is a geometric random variable with $\mathbb{P}\left(N_{ij}=n\right) = (1-p_{ij})^{n-1} p_{ij}$.
	
	We derive the mean and variance of $T_{ij}$ as follows (See Section 1.2 in the online supplement for the proof): 
	\begin{equation}
		\nonumber
		\mathbb{E}[T_{ij}] = \frac{\mathbb{E}[T_{i}]}{p_{ij}},
		\quad 
		\text{Var}[T_{ij}] = \frac{\text{Var}[T_i]}{p_{ij}} + (\mathbb{E} [T_i])^2 \frac{1-p_{ij}}{p_{ij}^2}.
	\end{equation}
	Thus, with the necessary notations defined in Table \ref{tab:not_thinning}, the mean and variance of the inter-arrival time of the thinned process are:
	\begin{equation}
		\frac{1}{\lambda_{ij}} = \frac{1}{\lambda_i p_{ij}},
		\quad 
		\sigma_{a,ij}^2 = \frac{\sigma_{a,i}^2}{p_{ij}} + \frac{1-p_{ij}}{\lambda_i^2 p_{ij}^2}.  
		\nonumber
	\end{equation}
	The uncertainty set of the inter-arrival times of the thinned process is:
	\begin{equation}
		\mathcal{U}^a_{ij}(n) = \left\{\mathbf{T}_{ij}(n) = (T_{ij}^1, T_{ij}^2, \dots, T_{ij}^n) \left| \frac{\sum_{l=k}^n T_{ij}^l - \frac{n-k+1}{\lambda_{ij}}}{\sqrt{n-k+1}} \geq -\Gamma_{a,ij}, \ \forall k, 1 \leq k \leq n \right\} \right. .
		\nonumber
	\end{equation}
	
	We can determine the variability parameter $\Gamma_{a,ij}$ with the derived mean $1/\lambda_{ij}$ and variance $\sigma_{a,ij}^2$ for the target probability level. The uncertainty set of the workloads of the thinned process is identical to the uncertainty set of workloads of application $i$ as follows:
	\begin{equation}
		\begin{gathered}
			\mathcal{U}^s_{ij}(n) = \mathcal{U}^s_{i}(n)
			= \left\{\mathbf{X}_i(n) = (X_i^1, X_i^2, \dots, X_i^n) \left| \frac{\sum_{l=k}^n X_i^l - 	\frac{n-k+1}{\mu_i}}{\sqrt{n-k+1}} \leq \Gamma_{s,i}, \ \forall k, 1 \leq k \leq n \right. \right. \\
			\left. \textrm{ and} \ \frac{\sum_{l=k}^{n-1} X_i^l - \frac{n-k}{\mu_i}}{\sqrt{n-k}} \leq \Gamma_{s,i}, \ \forall k, 1 \leq k \leq n-1 \right\}.
		\end{gathered}
	\end{equation}
	Similarly, we can determine the variability parameter $\Gamma_{s,i}$ with $\mu_{i}$ and $\sigma_{s,i}^2$ for the target probability level.
	
	\begin{table}[t]
		\caption{Notations for superposition of arrival processes}
		\label{tab:not_superpositioning}
		\centering
		\scriptsize
		\begin{tabular}{cc}
			\toprule
			Symbol & Description\\
			\midrule
			$\bar{\lambda}_j$ & Arrival rate for server $j$\\
			$\bar{\sigma}_{a,j}$ & Standard deviation of the inter-arrival time of arrival into server $j$\\
			$1/\bar{\mu}_j$ & Average workload for server $j$\\
			$\bar{\sigma}_{s,j}$ & Standard deviation of workload into server $j$\\
			\bottomrule
		\end{tabular}
	\end{table}
	
	\subsection{\emph{Superposition of arrival processes}}\label{subsec:superposition}
	
	Next, we aggregate multiple thinned processes to generate an actual aggregated arrival process to server $j$ and derive its uncertainty sets. After superposition, the dependence on inter-arrival times might arise and invalidate the CLT assumptions, causing issues with the uncertainty set. However, to keep the structure of the CLT-based uncertainty set, we apply the method from \citet{Bandi2015} for constructing the uncertainty set for the inter-arrival times, which includes all the instances of the aggregate process' inter-arrival times. For the workloads, in which dependence issue does not arise, we derive the exact moments of the aggregate process' workload random variable and directly construct the uncertainty set.
	
	First, we illustrate the set of inter-arrival times. In our model, all the uncertainty sets of inter-arrival times of the thinned processes to be merged are in CLT-based shape, that is, the term in the left-hand side of the inequality is inversely proportional to the square root of $n-k$ (or $n-k+1$). Therefore, we can apply the superposition operator in Theorem 5 of \citet{Bandi2015}. For a more general design of uncertainty sets considering the heavy-tailed distributions and generalized CLT, we refer to \citet{Bandi2015} for details. The uncertainty set of inter-arrival times, $\bar{\mathcal{U}}^a_j(n)$, for the server-wise aggregated process is as follows:
	\begin{equation}
		\begin{gathered}
			\bar{\mathcal{U}}^a_{j}(n) = \left\{\bar{\mathbf{T}}_{j}(n) = (\bar{T}_j^1, \bar{T}_j^2, \dots, \bar{T}_j^n) \left| \frac{\sum_{l=k}^n \bar{T}_{j}^l - \frac{n-k+1}{\bar{\lambda}_{j}}}{\sqrt{n-k+1}} \geq -\bar{\Gamma}_{a,j}, \ \forall k, 1 \leq k \leq n \right\} \right. .
			\label{eqn:usforsup}
		\end{gathered}
	\end{equation}
	This uncertainty set is slightly larger than the region where the aggregate process' inter-arrival times lie, as it considers the potential dependence between them.
	That is, the uncertainty set of the exact aggregate arrival intervals becomes the subset of the one in equation~\eqref{eqn:usforsup}. With necessary notations defined in Table~\ref{tab:not_superpositioning}, the mean arrival rate and variability parameter of the new uncertainty set are (See Section 1.3 in the online supplement for the proof):
	\begin{equation}
		\bar{\lambda}_j = \sum_{i \in \mathcal{A}_j} \lambda_{ij} = \sum_{i \in \mathcal{A}_j} \lambda_i p_{ij}, \quad	\bar{\Gamma}_{a,j} = \frac{\left(\sum_{i \in \mathcal{A}_j} \left( p_{ij}C_{a,i}^2 + (1-p_{ij})\right)\right)^{1/2} }{\sum_{i \in \mathcal{A}_j} \lambda_i p_{ij}},
		\label{eqn:params_arr}
	\end{equation}
	where $C_{a,i}^2$ is a squared coefficient of variation of the inter-arrival time of application $i$.
	
	Next, we describe the uncertainty set of workloads. From the results about the aggregated arrivals' workload in Section 7.1.2 of \citet{Gautam2012}, the mean and variance of the actual workload are:
	\begin{equation}
		\frac{1}{\bar{\mu}_j} = \frac{\sum_{i \in \mathcal{A}_j} \frac{\lambda_{ij}}{\mu_i}}{\sum_{i \in \mathcal{A}_j} \lambda_{ij}} = \frac{\sum_{i \in \mathcal{A}_j} p_{ij} \frac{\lambda_{i}}{\mu_i}}{\sum_{i \in \mathcal{A}_j} p_{ij} \lambda_{i}} , \quad
		\bar{\sigma}_{s,j}^2 = \frac{\sum_{i \in \mathcal{A}_j} \left(  \lambda_{ij} \left(\sigma_{s,i}^2 + \left(\frac{1}{\bar{\mu}_j} - \frac{1}{\mu_i}\right)^2\right) \right)  }{\sum_{i \in \mathcal{A}_j} \lambda_{ij}}.
		\label{eqn:ussup}
	\end{equation}
	With these exact moments of workload, the uncertainty set of service workloads, $\bar{\mathcal{U}}^a_j(n)$, for the server-wise aggregated process can be constructed as follows:
	\begin{equation}
		\begin{gathered}
			\bar{\mathcal{U}}^s_{j}(n) = \left\{\bar{\mathbf{X}}_j(n) = (\bar{X}_j^1, \bar{X}_j^2, \dots, \bar{X}_j^n)  \left| \frac{\sum_{l=k}^n \bar{X}_j^l - \frac{n-k+1}{\bar{\mu}_j}}{\sqrt{n-k+1}} \leq \bar{\Gamma}_{s,j}, \ \forall k, 1 \leq k \leq n \right. \right. \\
			\left. \textrm{ and } \ \frac{\sum_{l=k}^{n-1} \bar{X}_j^l - \frac{n-k}{\bar{\mu}_j}}{\sqrt{n-k}} \leq \bar{\Gamma}_{s,j}, \ \forall k, 1 \leq k \leq n-1 \right\} . \label{eqn:usforsup_workload}
		\end{gathered}
	\end{equation}
	As in equation~\eqref{eqn:functional_form_vp_PS}, the variability parameter for workloads is:
	\begin{equation}
		\begin{aligned}
			\bar{\Gamma}_{s,j} &= f_s\left(\bar{\mu}_j x_j, \bar{\lambda}_j, \bar{\sigma}_{a,j}, \frac{\bar{\sigma}_{s,j}}{x_j}; 1 - \varepsilon_j\right) .
		\end{aligned}
		\label{eqn:aggregated_gamma_s}
	\end{equation}
	Here, the workload mean and standard deviation are scaled according to the server speed because the derivation results in Section~\ref{subsec:main_variability_parameters} assume unit service rate.

	In the end, the uncertainty sets of the aggregated arrival process in equations~\eqref{eqn:usforsup} and \eqref{eqn:usforsup_workload} reflect both the characteristics of the external arrival process and routing policy. The variability parameters serve as a bridge between the external, thinned, and aggregated arrival processes. Thus, the relationship between the variability parameters of each process' uncertainty sets identifies the relationship between the processes via thinning and superposition.
	
	\subsection{\emph{Server-wise analysis}}
	\label{subsec:serverwise}
	
	With the uncertainty sets of the actual arrival process into server $j$ in equation~\eqref{eqn:usforsup}, QoS constraints in the optimization model can be expressed in terms of the server speed scaling. Using the results in equations~\eqref{eqn:params_arr}--\eqref{eqn:aggregated_gamma_s}, we construct a tractable condition that approximates the server $j$'s response time constraint, $\mathbb{P}(S_j \geq \delta_j) \leq \varepsilon_j$, similar to equation (\ref{eqn:slatoineq}) as follows:
	\begin{equation}
		\frac{ (\bar{\Gamma}_{a,j} + \bar{\Gamma}_{s,j})^2 \bar{\lambda}_j}{2\left(1-\frac{\bar{\lambda}_j}{\bar{\mu}_j x_j}\right)} + \frac{2-\frac{\bar{\lambda}_j}{\bar{\mu}_j x_j}}{\bar{\lambda}_j} \leq \delta_j. \label{eqn:slatoineq_servj}
	\end{equation}
	In equation~(\ref{eqn:slatoineq_servj}), $\rho$ is replaced by $\bar{\lambda}_j/\bar{\mu}_j x_j$ because we consider the scalable server speed instead of the unit speed at which unit workload is processed per unit time. The traffic intensity $\rho_j$ is the server-wise mean arrival rate $\bar{\lambda}_j$ multiplied by the scaled server-wise mean workload $1/\bar{\mu}_j x_j$. Additionally, we calculate the workload variability parameter $\bar{\Gamma}_{s,j}$ considering the server speed in equation~\eqref{eqn:slatoineq_servj}. There are no changes in the parameters related to the inter-arrival time.
	
	Based on the derivation results so far, we construct a new optimization model (M2) as follows:
	\begin{align}
		\textrm{(M2)} \quad \textrm{Min}\ &\sum_{j\in\mathcal{S}} \left( C_j(x_j) \frac{\sum_{i \in \mathcal{A}_j} p_{ij}\frac{\lambda_i}{\mu_i} }{x_j} + C_j\left(x^{min}_j\right) \left(1-\frac{\sum_{i \in \mathcal{A}_j}  p_{ij} \frac{\lambda_i}{\mu_i}}{x_j}\right)\right) \nonumber \\
		\textrm{s.t.}\ & 0 \leq p_{ij} \leq 1, \quad\forall i \in \mathcal{A}, \forall j \in \mathcal{S}_i, \label{m2const:prob} \\
		& \sum_{j \in \mathcal{S}_i} p_{ij} = 1, \quad\forall i \in \mathcal{A}, \label{m2const:prob2}\\
		& x_j^{min} \leq x_j \leq x_j^{max},\quad\forall j\in\mathcal{S} \label{m2const:speed},\\
		& \bar{\lambda}_j = \sum_{i \in \mathcal{A}_j} \lambda_{ij} = \sum_{i \in \mathcal{A}_j} \lambda_i p_{ij} , \quad \forall j \in \mathcal{S},\label{m2const:uamean}\\
		&\bar{\sigma}_{a,j}^2 = \bar{\Gamma}_{a,j}^2 = \frac{\sum_{i \in \mathcal{A}_j} \left( p_{ij}C_{a,i}^2 + (1-p_{ij})\right)}{\left(\sum_{i \in \mathcal{A}_j} \lambda_i p_{ij}\right)^2}, \quad \forall j \in \mathcal{S}, \label{m2const:gamma_a_function} \\
		& \frac{1}{\bar{\mu}_j} =  \frac{\sum_{i \in \mathcal{A}_j} p_{ij} \frac{\lambda_i}{\mu_i}}{\sum_{i \in \mathcal{A}_j} p_{ij} \lambda_{i}}, \quad \forall j \in \mathcal{S}, \label{m2const:usmean}\\
		&\bar{\sigma}_{s,j}^2 = \frac{\sum_{i \in \mathcal{A}_j} \lambda_{ij} \Big(\sigma_{s,i}^2 + \Big(\frac{1}{\bar{\mu}_j} - \frac{1}{\mu_i}\Big)^2\Big)}{\sum_{i \in \mathcal{A}_j} \lambda_{ij}}, \quad \forall j \in \mathcal{S}, \label{m2const:ussig}\\
		& \bar{\Gamma}_{s,j} =  f_s\left(\bar{\mu}_j x_j, \bar{\lambda}_j, \bar{\sigma}_{a,j}, \frac{\bar{\sigma}_{s,j}}{x_j}; 1 - \varepsilon_j\right), \quad \forall j \in \mathcal{S}, \label{m2const:gamma_s_function}\\
		&\frac{ (\bar{\Gamma}_{a,j} + \bar{\Gamma}_{s,j})^2 \bar{\lambda}_j}{2\left(1-\frac{\bar{\lambda}_j}{\bar{\mu}_j x_j}\right)} + \frac{2-\frac{\bar{\lambda}_j}{\bar{\mu}_j x_j}}{\bar{\lambda}_j} \leq \delta_j, \quad \forall j \in \mathcal{S}. \label{m2const:sla}
	\end{align}
	
	The objective function of (M2) is the same as that of (M1). Constraints (\ref{m2const:prob})--(\ref{m2const:speed}) are the same as constraints (\ref{const:1})--(\ref{const:3}) in the optimization model (M1). The rest of the constraints in the model (M2) are the approximate condition of the probabilistic constraint (\ref{const:4}) in the model (M1). Constraints (\ref{m2const:uamean}) and \eqref{m2const:gamma_a_function} are identical to the result in equation (\ref{eqn:params_arr}), which denotes the mean and variance of the inter-arrival time of aggregated arrival process into server $j$. Constraints~(\ref{m2const:usmean}) and (\ref{m2const:ussig}) are the results from equation (\ref{eqn:ussup}), which denotes the mean and variance of the workload (job size) of aggregated arrival process into server $j$. Constraints \eqref{m2const:gamma_a_function} and \eqref{m2const:gamma_s_function} determine the variability parameters to fit the size of uncertainty sets to target probability level $1-\varepsilon_j$ by the method explained in Section~\ref{subsec:main_variability_parameters}. Constraint~\eqref{m2const:sla} bounds the approximate sojourn time quantile as in equation~\eqref{eqn:slatoineq_servj}. 
	
	\begin{table}[t!]
		\caption{Arrival process setting}
		\label{tab:not_arrival}
		\centering
		\resizebox{0.8\textwidth}{!}{%
			\begin{tabular}{@{}cccccccc@{}}
				\toprule
				\multirow{2}{*}{Application type ($i$)} & \multicolumn{3}{c}{Inter-arrival time random variable} & \multicolumn{3}{c}{Workload random variable} &\multirow{2}{*}{Instant demand rate}\\ \cmidrule(l){2-7} 
				& Distribution      & Mean ($1/\lambda_i$)     & SCOV     & Distribution  & Mean ($1/\mu_i$)  & SCOV  \\ \midrule
				1                                & Lognormal         & 0.25     & 2        & Lognormal     & 5     & 1.5   & 20\\ \midrule
				2                                & Lognormal         & 0.5      & 1.5      & Lognormal     & 10    & 2     & 20\\ \midrule
				3                                & Exponential       & 0.25     & 1        & Lognormal     & 5     & 1     & 20\\ \midrule
				4                                & Lognormal         & 0.1      & 0.8      & Lognormal     & 2     & 0.8   & 20\\ \midrule
				5                                & Lognormal         & 0.2      & 2        & Lognormal     & 3     & 0.5   & 15\\ \bottomrule
			\end{tabular}
		}
	\end{table}

	\section{Numerical experiments}	\label{sec:exp}
	
	Here, we show the numerical results of applying the static policy derived through solving the model (M2) in Section~\ref{sec:sla}. As mentioned in Section \ref{sec:introduction}, most of the operation-level dynamic control algorithms for data centers only targeted routing or scaling. A few considering both simultaneously narrowly considered the sojourn time-related probabilistic QoS constraints. Hence, we select the dynamic speed scaling and load balancing algorithm in \citet{Ko2014} as a benchmark model. For a meaningful comparison, we design our experimental settings similar to those in \citet{Ko2014}. The experimental details, including solving the optimization model (M2) and running a simulation, are explained in Section 5 in the online supplement.
	
	Section~\ref{subsec:comparedynamic} explains the simulation results of the proposed static policy and the reference dynamic control. Additionally, we compare the power consumption and violation probability by changing the QoS-related parameters, $\delta$ and $\varepsilon$. Section~\ref{subsec:discussion_static} provides discussions about the proposed static control. Section~\ref{subsubsec:discussion_relative_performance} discusses the circumstance under which the static policy is relatively appealing and draws insights from the comparison with the dynamic approach. Section~\ref{subsubsec:additional_discussions} discusses the feasibility and scalability of the proposed method.

	\subsection{\emph{Comparison with dynamic control}}
	\label{subsec:comparedynamic}	
	
	\begin{table}[t!]
		\caption{Server setting}
		\label{tab:not_server}
		\centering
		\scriptsize
		\resizebox{0.5\textwidth}{!}{
		\begin{tabular}{@{}ccccccccc@{}}
			\toprule
			Server ($j$)  & $x^{min}_j$  &  $x^{max}_j$&  $K_j$ &  $\alpha_j$ & $n_j$  &  $\delta_j$ &  $\varepsilon_j$ & $\mathcal{A}_j$ \\ \midrule
			\multicolumn{1}{c}{1} & \multicolumn{1}{c}{5} & \multicolumn{1}{c}{100} & \multicolumn{1}{c}{150} & \multicolumn{1}{c}{1/3} & \multicolumn{1}{c}{3} & \multicolumn{1}{c}{$\delta$} & \multicolumn{1}{c}{$\varepsilon$} & \multicolumn{1}{c}{1}     \\ 
			\multicolumn{1}{c}{2} & \multicolumn{1}{c}{7} & \multicolumn{1}{c}{102} & \multicolumn{1}{c}{250} & \multicolumn{1}{c}{0.2} & \multicolumn{1}{c}{3} & \multicolumn{1}{c}{$\delta$} & \multicolumn{1}{c}{$\varepsilon$} & \multicolumn{1}{c}{1}     \\ 
			\multicolumn{1}{c}{3} & \multicolumn{1}{c}{6} & \multicolumn{1}{c}{99}  & \multicolumn{1}{c}{220} & \multicolumn{1}{c}{1}   & \multicolumn{1}{c}{3} & \multicolumn{1}{c}{$\delta$} & \multicolumn{1}{c}{$\varepsilon$} & \multicolumn{1}{c}{1,2}   \\ 
			\multicolumn{1}{c}{4} & \multicolumn{1}{c}{5} & \multicolumn{1}{c}{105} & \multicolumn{1}{c}{150} & \multicolumn{1}{c}{2/3} & \multicolumn{1}{c}{3} & \multicolumn{1}{c}{$\delta$} & \multicolumn{1}{c}{$\varepsilon$} & \multicolumn{1}{c}{1,2,3} \\ 
			\multicolumn{1}{c}{5} & \multicolumn{1}{c}{7} & \multicolumn{1}{c}{100} & \multicolumn{1}{c}{300} & \multicolumn{1}{c}{0.8} & \multicolumn{1}{c}{3} & \multicolumn{1}{c}{$\delta$} & \multicolumn{1}{c}{$\varepsilon$} & \multicolumn{1}{c}{2,3}   \\ 
			\multicolumn{1}{c}{6} & \multicolumn{1}{c}{8} & \multicolumn{1}{c}{102} & \multicolumn{1}{c}{350} & \multicolumn{1}{c}{0.4} & \multicolumn{1}{c}{3} & \multicolumn{1}{c}{$\delta$} & \multicolumn{1}{c}{$\varepsilon$} & \multicolumn{1}{c}{2,3}   \\ 
			\multicolumn{1}{c}{7} & \multicolumn{1}{c}{6} & \multicolumn{1}{c}{100} & \multicolumn{1}{c}{220} & \multicolumn{1}{c}{3/7} & \multicolumn{1}{c}{3} & \multicolumn{1}{c}{$\delta$} & \multicolumn{1}{c}{$\varepsilon$} & \multicolumn{1}{c}{3}     \\ 
			\multicolumn{1}{c}{8} & \multicolumn{1}{c}{7} & \multicolumn{1}{c}{105} & \multicolumn{1}{c}{350} & \multicolumn{1}{c}{0.5} & \multicolumn{1}{c}{3} & \multicolumn{1}{c}{$\delta$} & \multicolumn{1}{c}{$\varepsilon$} & \multicolumn{1}{c}{4,5}   \\ 
			\multicolumn{1}{c}{9} & \multicolumn{1}{c}{8} & \multicolumn{1}{c}{102} & \multicolumn{1}{c}{400} & \multicolumn{1}{c}{0.6} & \multicolumn{1}{c}{3} & \multicolumn{1}{c}{$\delta$} & \multicolumn{1}{c}{$\varepsilon$} & \multicolumn{1}{c}{4,5}   \\ 
			\multicolumn{1}{c}{10} & \multicolumn{1}{c}{10} & \multicolumn{1}{c}{105} & \multicolumn{1}{c}{700} & \multicolumn{1}{c}{4/9} & \multicolumn{1}{c}{3} & \multicolumn{1}{c}{$\delta$} & \multicolumn{1}{c}{$\varepsilon$} & \multicolumn{1}{c}{5}   \\ \bottomrule
		\end{tabular}
		}
		\\
		Power function : $C_j(x_j) = K_j + \alpha_j x_j^{n_j}$, QoS constraint : $\mathbb{P}(S_j \geq \delta_j) \leq \varepsilon_j$\\
	\end{table}
	
	We consider a system with five applications and ten servers as \citet{Ko2014}. For illustrative purposes, we use a small-size problem for comparison and discuss the scalability issue of solving a large-size problem later in Section~\ref{subsubsec:additional_discussions}. The settings for the arrival processes of each application and server are listed in Tables~\ref{tab:not_arrival} and \ref{tab:not_server}, respectively. SCOV means the squared coefficient of variation.
	
	The arrival processes of the applications are independent renewal processes consisting of inter-arrival times and workloads following general distributions. We assume that the servers' power functions are cubic functions ($C_j(x_j) = K_j + \alpha_j x_j^3$). Each server can have its speed ranges ($[x^{min}_j, x^{max}_j]$), parameters of the power function ($K_j, \alpha_j$), and a set of applications it can handle ($\mathcal{A}_j$). For simplicity, we use the same $\delta=\delta_j$ and $\varepsilon=\varepsilon_j$ for all $j \in \mathcal{S}$ as in \citet{Ko2014}; note that having different $\delta_j$ and $\varepsilon_j$ does not affect the computational complexity. The routing probability matrix and the server speeds of the static policy are obtained by solving the optimization model (M2) with an open-source optimization solver IpOpt (version 0.7.0).
	\begin{table}[h!]
		\caption{Comparison result for $\delta=5$ and $\varepsilon = 0.05$}
		\label{tab:not_comparison_5_005}
		\centering
		\resizebox{0.9\textwidth}{!}{%
			\begin{tabular}{@{}ccccccccccccc@{}}
				\toprule
				\multirow{3}{*}{\begin{tabular}[c]{@{}c@{}}Control\\ Type\end{tabular}} &
				\multirow{3}{*}{\begin{tabular}[c]{@{}c@{}}Average Power \\ Consumption\\ (per unit time)\end{tabular}} &
				\multicolumn{11}{c}{Server $j$} \\ \cline{3-13} 
				&
				&
				\multirow{2}{*}{$j$} &
				\multirow{2}{*}{1} &
				\multirow{2}{*}{2} &
				\multirow{2}{*}{3} &
				\multirow{2}{*}{4} &
				\multirow{2}{*}{5} &
				\multirow{2}{*}{6} &
				\multirow{2}{*}{7} &
				\multirow{2}{*}{8} &
				\multirow{2}{*}{9} &
				\multirow{2}{*}{10} \\
				&                           &   &        &        &         &        &        &         &        &        &        &        \\ \midrule
				\multirow{2}{*}{Dynamic} & \multirow{2}{*}{18421.07} & $\mathbb{E}[x_j]$ & 9.824  & 13.054 & 8.220   & 10.691 & 9.779  & 14.499  & 11.458 & 13.544 & 12.372 & 12.604 \\
				&                           & $\mathbb{P}(S_j \geq 5)$ & 0      & 0      & 0.00027 & 0      & 0      & 0.00007 & 0      & 0      & 0      & 0      \\ \midrule
				\multirow{2}{*}{Static}  & \multirow{2}{*}{16767.49} & $\mathbb{E}[x_j]$ & 13.834 & 17.324 & 18.903 & 17.126  & 15.313 & 25.875  & 15.382 & 14.297 & 13.050 & 15.291 \\
				&                           & $\mathbb{P}(S_j \geq 5)$ & 0.0246 & 0.0248 & 0.0075  & 0.0061 & 0.0113 & 0.0098  & 0.0407 & 0.0118 & 0.0118 & 0.0409 \\ \midrule
			\end{tabular}%
		}
	\end{table}
	
	First, we take a close look at a specific operation example under $\delta = 5$ and $\varepsilon = 0.05$, or the QoS constraints being $\mathbb{P}(S_j \geq 5) \leq 0.05, \ \forall j \in \mathcal{S}$. Table~\ref{tab:not_comparison_5_005} summarizes the experimental results. The average power consumption per unit time of the static policy is approximately $9\%$ lower than that of the dynamic algorithm. The dynamic algorithm shows violation probabilities far smaller than the target value $\varepsilon = 0.05$, whereas the proposed static control shows violation probabilities that are not extremely low. It might seem counterintuitive that the mean server speed, $\mathbb{E}[x_j]$, is generally higher under the static control policy, which has lower average power consumption. However, this phenomenon can be explained by considering the characteristics of the control policies and the power function used in the study. The power function is assumed to be cubic, meaning that power consumption increases much more rapidly as server speed increases. Static control operates servers at a constant speed regardless of workload (except for idle time), while dynamic control adjusts the speed of the servers based on the remaining workload. Dynamic control is generally more energy efficient during low-traffic periods. However, when traffic suddenly increases, it might operate at a higher speed for some time to meet QoS constraints, leading to higher energy consumption. Contrastingly, static control will have a manageable backlog of workloads, because it always processes services at a fixed speed, if not idle. As a result, static control might lead to lower average energy consumption, related to the third-order moment of server speed, although it has a higher average operating speed, the first-order moment, compared to dynamic control, particularly in environments with weaker QoS constraints.
	
	\begin{figure*}[t!]
		\centering
		\includegraphics[width=0.9\linewidth]{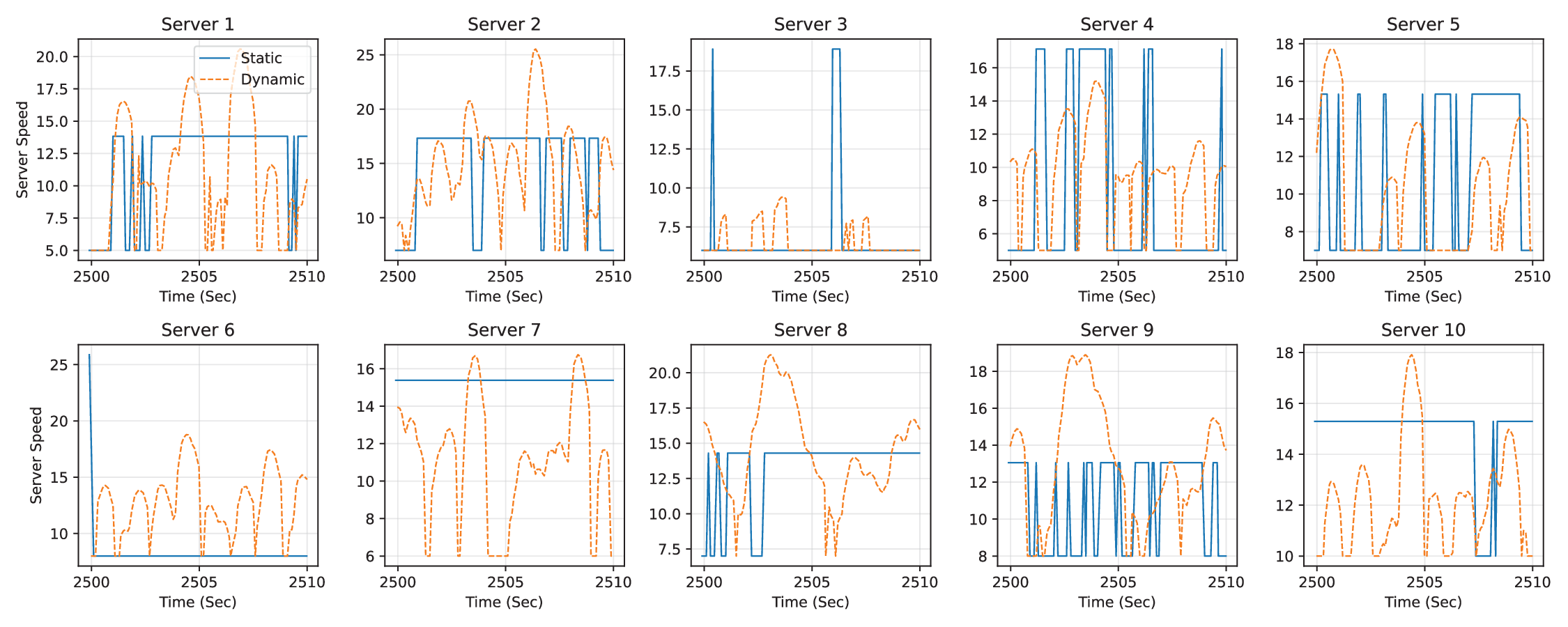}
		\caption{Server speeds for $\delta =5$ and $\varepsilon = 0.05$}
		\label{fig:speeds_5_005}
	\end{figure*}
	
	Figure~\ref{fig:speeds_5_005} shows a part of the server speeds from the simulation result. The dashed and solid lines denote the server speeds from the dynamic and static control over time, respectively. Under the proposed static policy, servers run at the predetermined speed derived from the optimization model only when unfinished jobs are left and run at the lowest possible speed $x^{min}_j$ when the server $j$ is idle. Thus, the server speed under static policy appears like a step-wise function. However, the server speed under dynamic control appears nearly continuous because it changes according to the remaining workload at the moment. In summary, the server speed in the dynamic control continuously varies, whereas the static control speed switches only between two values similar to a bang-bang type control.
	
	\begin{figure*}[t]
		\centering
		\includegraphics[width=0.9\linewidth]{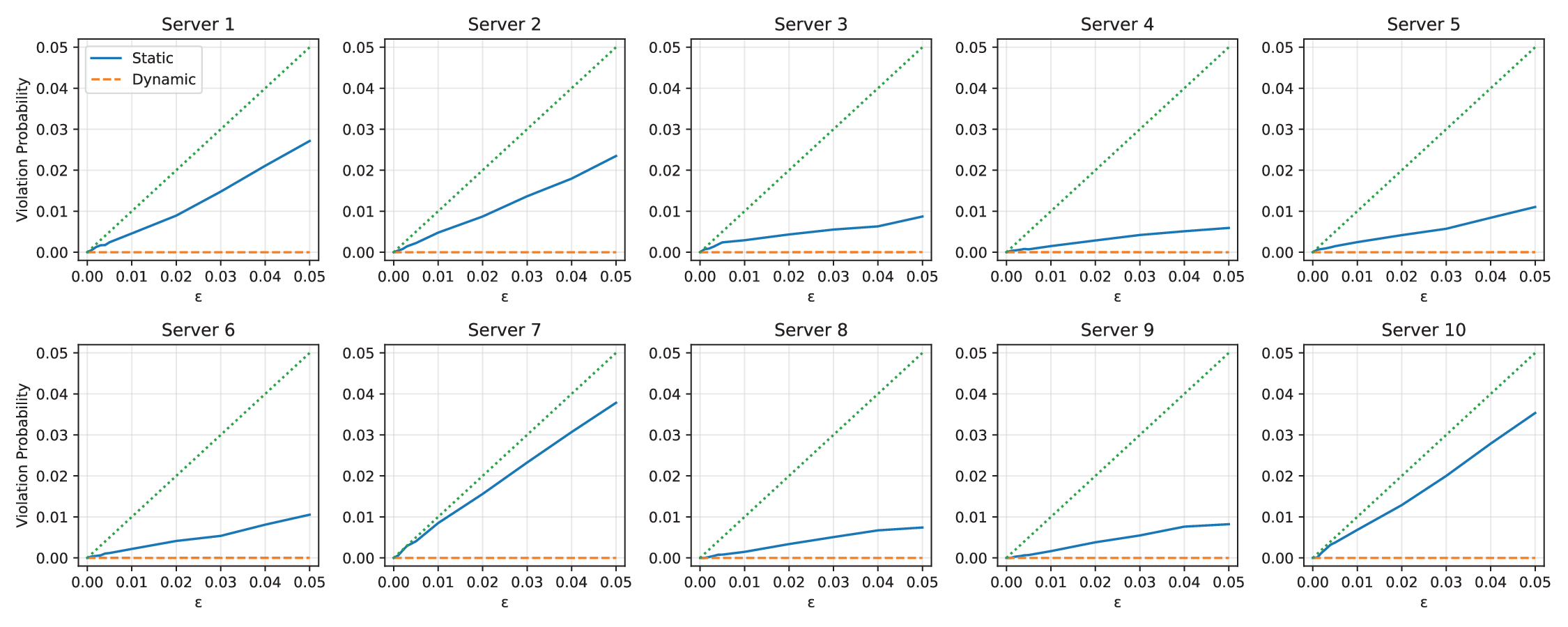}
		\caption{Violation probabilities for $\delta =5$}
		\label{fig:slas}
	\end{figure*}
	
	Next, we examine the violation probabilities under the dynamic and static policies with different target probabilities $\varepsilon$'s and the same $\delta = 5$. The results are plotted in Figure~\ref{fig:slas}. Both policies successfully satisfy QoS conditions; however, the result of the dynamic control is overly conservative, particularly with larger $\varepsilon$, that is, the dashed lines denoting the violation probabilities of the dynamic control fall below the dotted lines denoting the target value $\varepsilon$ but significantly close to zero. The solid lines denoting the result of the static policy, however, indicate an adequate level of violation probability while satisfying QoS conditions.
	
\begin{figure}[t]
		\centering
		\includegraphics[width=\linewidth]{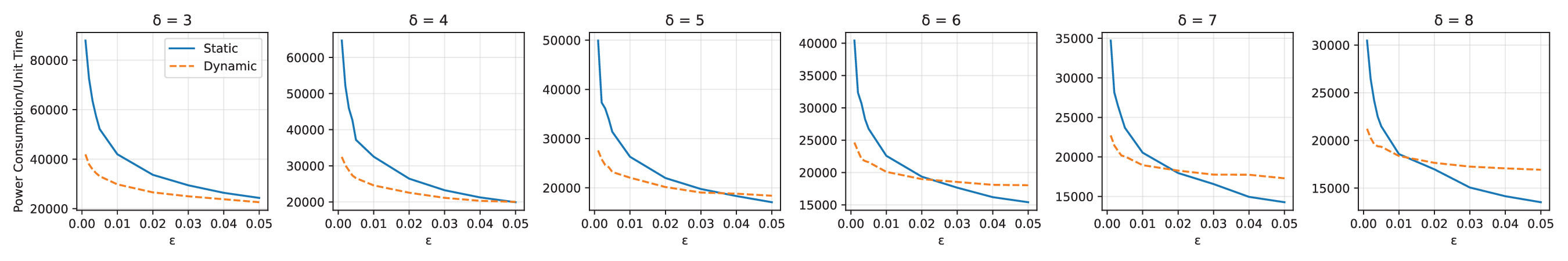}
		\caption{Average power consumptions for $\delta=3 $ to $8$}
		\label{fig:averagepowers}
	\end{figure}

	Finally, we compare the energy consumption of both policies as $\delta$ and $\varepsilon$ vary. Figure~\ref{fig:averagepowers} shows how the average power consumption changes as $\varepsilon$ grows for a specific $\delta$ under dynamic (dashed lines) and static (solid lines) policies. As the response time threshold $\delta$ or the violation probability bound $\varepsilon$ increases, the relative performance of the static policy improves. We expected this result because our construction of a tractable condition that approximates the QoS constraint is tighter than the heavy-traffic-based approach. For the detailed discussion about the sojourn time quantile estimation, we refer the readers to Section 2 in the online supplement. Thus, we derive a tighter but approximative sojourn time upper bound by using the concepts of the CLT and uncertainty sets. The fundamental difference of the operation policy, whether dynamic or static, causes the dynamic policy to outperform under the strict QoS conditions. However, under the loose QoS conditions, the benefits of our robust queueing theoretic approach appears to overcome the structural limitation of static operation. When $\varepsilon = 0.05$ and $\delta = 8$, the static policy consumes approximately $73\%$ of the energy of the dynamic one while successfully satisfying the QoS conditions.
	
	\subsection{\emph{Discussion on static control}}
	\label{subsec:discussion_static}
	
	This section assesses the effectiveness of the proposed static control approach from various angles. Section~\ref{subsubsec:discussion_relative_performance} examines the situations in which the static policy is particularly appealing and provides insights based on a comparison with the dynamic approach. Section~\ref{subsubsec:additional_discussions} explores the feasibility and scalability of the proposed method for deriving the static policy.
	
	\subsubsection{\emph{Discussion on relative performance}}
	\label{subsubsec:discussion_relative_performance}
	
	The conservative result in the violation probability described in Section~\ref{subsec:comparedynamic} originates from how tightly the QoS conditions are satisfied. Mathematically, the aforementioned phenomenon is related to the derivation of the approximate conditions for QoS constraints. Under the strict QoS conditions---that is, small $\varepsilon$ and $\delta$ values---the traffic intensity may not be heavy enough to apply the heavy-traffic approximations. Furthermore, \citet{Ko2014} used a loose upper bound of the heavy-traffic result to make the constraint linear, making the result more conservative even under the weak QoS conditions. However, the robust queueing theory does not require assumptions on traffic intensity to construct uncertainty sets, and the proposed static policy takes advantage of it.
	
	We also investigate the effect of different factors, {more than} $\varepsilon$ and $\delta$, on the relative performance of the static policy. Because of space limitation, we refer the reader to Section 4 in the online supplement for the experimental details, and we only discuss the results here.
	
	First, the dynamic policy performs relatively better as the arrival rate increases. We can interpret that a higher arrival rate makes traffic intensity closer to one, and consequently, the heavy-traffic approximation of the dynamic control becomes more accurate. Addition, as the variance of the inter-arrival time distribution of the external arrival process increases, the static policy performs relatively better. A larger variance of the inter-arrival time causes a larger variance in the system's status, consequently increasing server speeds to meet the QoS conditions. This variability causes the traffic intensity and accuracy of the heavy-traffic approximation of the dynamic policy to decrease, while the static control can respond resiliently to varying $\rho$.
	
	Second, as the power functions of servers become more heterogeneous, the static policy performs relatively better. Generally, the performance of the static policy is known to degrade as heterogeneity within a system grows \citep{Chen2015}. We, however, find that the way of handling the QoS conditions can produce this counter-intuitive result. We observe that service requests are routed to more energy-efficient servers. If the heterogeneity in the power functions increases, this tendency becomes more apparent; there are busy servers and idle servers. The accuracy of the heavy-traffic approximation varies among servers (that is, accurate for servers in heavy traffic and inaccurate with low traffic). The robust queueing theory in the static policy expects consistent accuracy because of its insensitivity to the traffic intensity.	
	
	\subsubsection{\emph{Additional discussions}}
	\label{subsubsec:additional_discussions}
	
	In this section, we discuss the feasibility and scalability of the proposed method. We provide experimental results with our interpretations of them.
	
	For the feasibility of the problem, we successfully obtain the robust control policy under the setting explained in Section~\ref{subsec:comparedynamic}, where $\varepsilon$ and $\delta$ values are in the ranges $[0.001,0.05]$ and $[3,8]$, respectively. In addition, via numerical experiments, we confirm that we could obtain the optimal server speeds and routing probability matrix for a fairly large range of parameters (i.e., $\varepsilon$ and $\delta$ seem to practically have no restriction). Even for the strict conditions with $\varepsilon = 10^{-5}$ or $\delta = 1$, we successfully obtain the optimal solution of (M2).
	
\begin{figure}
		\begin{subfigure}{0.49\textwidth}
			\centering
			\includegraphics[width=0.45\linewidth]{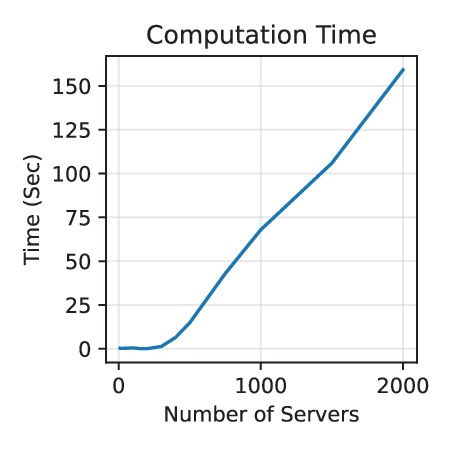}
			\caption{Computation time under different system scales}
			\label{fig:scalability_computimes}
		\end{subfigure}
		\begin{subfigure}{0.49\textwidth}
			\centering
			\includegraphics[width=0.9\linewidth]{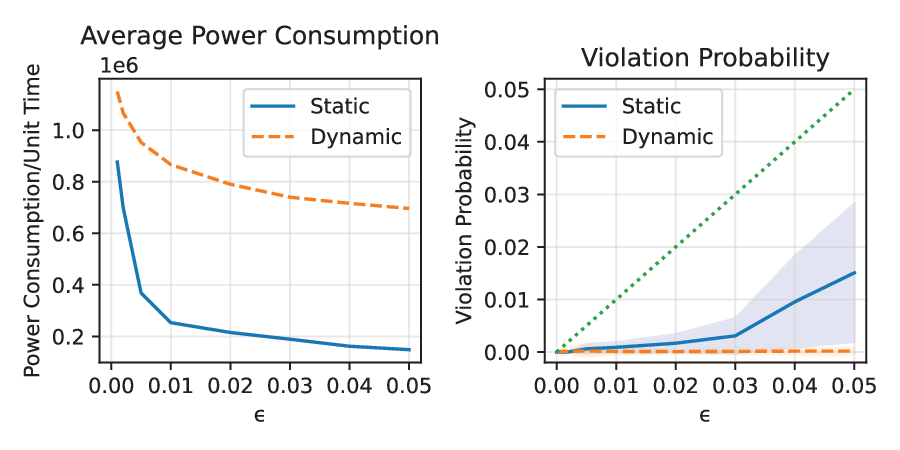}
			\caption{Comparison results for a system with 100 servers}
			\label{fig:scalability_energy_viol_probs}
		\end{subfigure}
		\caption{Experiment results for scalability analysis}
	\end{figure}

	For the scalability, the dynamic algorithm in \citet{Ko2014} requires only real-time monitoring of servers and communications between servers and a router. There are practically no scalability issues because it is a distributed algorithm. However, the proposed static policy requires solving the optimization problem (M2). We conduct numerical experiments to confirm whether the proposed methodology is applicable when the system size is large. We measure the computation time to obtain the static policy by solving the optimization problem (M2) with different system sizes. Because of space limitation, we defer the related experimental details to Section 3 in the online supplement and briefly discuss the result here. We successfully obtain the optimal solution of the optimization model and the corresponding static policy, even for a large system scale. Figure~\ref{fig:scalability_computimes} shows the computation time according to the number of servers with QoS condition parameters $\delta = 8,\  \varepsilon = 0.05$. The optimization problem for the system with 2,000 servers can be solved within 3 minutes. Figure~\ref{fig:scalability_energy_viol_probs} depicts the average power consumption per unit time and the mean ($\pm$ standard deviation) of violation probabilities of all servers for the system with $100$ servers regarding the QoS conditions with $(\delta, \ \varepsilon) = (8, \ 0.001)$. The result shows that the static policy outperforms the dynamic control for almost every $\varepsilon$ because of the increased variability of the arrival process. Additionally, we observe that the violation probabilities are not extremely low but still below threshold $\varepsilon$.
		
	\section{Conclusion}
	\label{sec:con}
	
	In this study, we proposed a methodology to derive an energy-efficient static policy while maintaining a satisfactory level of service. Each server was modeled as a $G/G/1/PS$ queue, and QoS conditions restricted the tail probabilities of response times in the mathematical model. The proposed approach based on the robust queueing theory used uncertainty sets to define the domain of stochastic primitives: inter-arrival times and workloads.
	
	We derived an approximative upper bound of sojourn times from the uncertainty sets and proposed a method to determine the size of the uncertainty sets, that is, variability parameters. Numerical experiments confirmed that the sojourn time bound accurately approximates the probabilistic quantile of steady-state sojourn time. We constructed a tractable inequality that replaces the QoS constraint, which was initially a probabilistic statement. The optimal solution of the reformulated optimization model deduced a static policy consisting of routing probability matrix and server speeds. 
	
	The results of the numerical experiments show that under the weak QoS conditions with a generous response time threshold and violation probability level, the proposed static policy outperformed the dynamic algorithm of \citet{Ko2014} in energy consumption while satisfying the QoS conditions. Moreover, the proposed static policy is practically scalable to {relatively} large systems.
	
	In future research, the current model could be refined by including the concept of multi-tier architecture and multi-core processors to represent a more realistic system. Additionally, we could consider an alternative design for uncertainty sets in different disciplines (e.g., last come first served and shortest remaining time first). Subsequently, the analysis procedure for various performance measures could also be devised. We could extend the static control policy proposed in this study to consider time-varying arrival processes. A dynamic control algorithm based on {the} robust queueing theory that utilizes real-time information in constructing uncertainty sets could also be investigated.
	
	\if0\blind{
	\section*{Acknowledgements}
	We would like to thank the Editor, Department Editor, Associate Editor and anonymous reviewers for their insightful comments to strengthen this paper. This work was supported in part by the National Research Foundation of Korea (NRF) grant funded by the Korea government (MSIT) (No. NRF-2021R1A2C1094699 and NRF-2021R1A4A1031019) and in part by Korea Institute for Advancement of Technology (KIAT) grant funded by the Korea Government (MOTIE) (P0008691, HRD Program for Industrial Innovation).
	} \fi
	
	\if0\blind{
	\section*{Notes on contributors}
	\textit{Seung Min Baik} received the B.S. degree in industrial and management engineering from the Pohang University of Science and Technology (POSTECH), Pohang, South Korea, in 2016, where he is currently pursuing the Ph.D. degree in industrial and management engineering. His research interests include robust decision-making in stochastic systems and reliability engineering in stochastic simulation models.
	
	\noindent \textit{Young Myoung Ko} is an associate professor in the Department of Industrial and Management Engineering at Pohang University of Science and Technology (POSTECH), Pohang, South Korea. He received the B.S. and M.S. degrees in industrial engineering from Seoul National University, Seoul, South Korea, and the Ph.D. degree in industrial engineering from Texas A \& M University, College Station, TX, USA, in 1998, 2000, and 2011, respectively. His research interests include, but are not limited to, simulation and optimization of stochastic systems, such as telecommunication networks, ICT infrastructure, and renewable energy systems.
	} \fi
		
	
	\spacingset{1}

\end{document}